\documentclass[letterpaper,11pt]{article}

\usepackage{dynkin-diagrams}
\usepackage{cite}
\usepackage{amsmath}
\usepackage{amssymb}
\usepackage{amsthm}
\usepackage{mathrsfs}
\usepackage{xypic}
\usepackage{tikz-cd}
\usepackage{chngcntr}

\pgfkeys{/Dynkin diagram, edge length=1cm,
fold radius=.6cm, 
root-radius=.11cm,
indefinite edge/.style={
draw=black, fill=white, dotted,
thin}}

\usepackage{hyperref}
\usepackage{color}
\definecolor{darkred}{rgb}{0.65,0.15,0}
\hypersetup{pdfborder={0 0 0},colorlinks=true,urlcolor=darkred,citecolor=blue,linkcolor=darkred,linktocpage=true}

\usepackage{float} 

\usepackage{geometry}

\newgeometry{vmargin={35mm}, hmargin={35mm}} 

\usepackage{keyval}
\makeatletter
\define@key{setpar}{left}[0pt]{\leftmargin=#1}
\define@key{setpar}{right}[0pt]{\rightmargin=#1}
\define@key{setpar}{both}{\leftmargin=#1\relax\rightmargin=#1}
\makeatother

\makeatletter
\renewcommand*\env@matrix[1][\arraystretch]{%
  \edef\arraystretch{#1}%
  \hskip -\arraycolsep
  \let\@ifnextchar\new@ifnextchar
  \array{*\c@MaxMatrixCols c}}
\makeatother  

\def\eg{{\it e.g.}}
\def\ie{{\it i.e.}}

\def\adj{\hbox{\bf adj}}

\def\End{\mathrm{End}}

\def\DWeight#1#2#3{\bigl(\raise2.5pt\hbox{${}_{#1}$}{}^{#2}_{#3}\bigr)}

\def\AAWeight#1#2{\bigl(\raise0pt\hbox{${}^{#1}_{#2}$}\bigr)}

\def\fg{{\mathfrak g}}

\def\sl{{\mathfrak{sl}}}

\def\nn{\nonumber}

\def\gl{\mathfrak{gl}}

\def\*{\partial}

\def\BB{{\mathscr B}}

\def\RR{{\mathbb R}}
\def\CC{{\mathbb C}}
\def\ZZ{{\mathbb Z}}

\def\mA{{\mathscr A}}

\newcommand{\dd}{{\mathsf{d}}}
\newcommand{\KK}{{\mathsf{K}}}

\def\compl{{\scriptscriptstyle\complement}}

\numberwithin{equation}{section}

\newlength\symlength
\newlength\pluslength
\DeclareRobustCommand\lplus{
 \ensuremath{
  \mathop{
   \begin{tikzpicture}[line width=0.1ex]
    \useasboundingbox (-1ex, -0.7ex) rectangle (.7ex, 1ex);
\draw (55:\symlength) arc (55:305:\symlength);
\draw (-\pluslength,0) -- (\pluslength,0);
\draw (0,-\pluslength) -- (0,\pluslength);
   \end{tikzpicture}
  }\nolimits
 }
}

\def\inplus{\lplus}

\def\fa{\mathfrak{a}}

\def\Hom{\mathrm{Hom}}
\def\End{\mathrm{End}}

\def\Tn#1{\hskip.2pt{\buildrel{\scriptscriptstyle #1}\over T}\hskip.2pt}
\def\Jn#1{\hskip.2pt{\buildrel{\scriptscriptstyle \,#1\!}\over J}\hskip.2pt}
\def\En#1{\hskip.2pt{\buildrel{\scriptscriptstyle \,#1\!}\over E}\hskip.2pt}

\makeatletter
\newcommand{\oset}[3][0ex]{%
  \mathrel{\mathop{#3}\limits^{
    \vbox to#1{\kern-3\ex@
    \hbox{$\scriptscriptstyle#2$}\vss}}}}
\makeatother

\allowdisplaybreaks

\begin{document}


\frenchspacing

\null\vspace{-28mm}

\includegraphics[height=2cm]{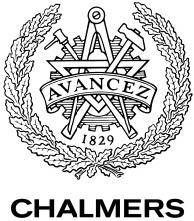}
\hspace{2mm}

\vspace{-12mm}
{\flushright Gothenburg preprint \\ 
}

\vspace{4mm}

\hrule

\vspace{16mm}


\thispagestyle{empty}

\begin{center}
  {\Large \bf \sc Gradient structures from extensions}
  \\[3mm]
  {\Large \bf \sc of over-extended Kac--Moody algebras}
    \\[10mm]
    
{\large
Martin Cederwall${}^1$ and Jakob Palmkvist${}^{2}$}

\vspace{10mm}
       {\footnotesize ${}^1${\it Department of Physics,
         Chalmers Univ. of Technology,\\
 SE-412 96 Gothenburg, Sweden}}

\vspace{2mm}
       {\footnotesize ${}^2${\it Department of Mathematical Sciences,
        Chalmers Univ. of Technology,\\
 SE-412 96 Gothenburg, Sweden}}

\end{center}

\vfill

\begin{quote}
  
\textbf{Abstract:} 
Over-extended Kac--Moody algebras contain so-called gradient structures---a $\gl(d)$-covariant level decomposition of the algebra contains strings of modules at different levels that can be interpreted as spatial gradients.
We present an algebraic origin for this phenomenon, based on the recently introduced Lie algebra extension
of an over-extended Kac--Moody algebra by its fundamental module \cite{Cederwall:2021ymp}, appearing in tensor hierarchy algebra super-extensions of over-extended Kac--Moody algebras.
The extensions are described in terms of Lie algebra cohomology, vanishing for finite-dimensional simple Lie algebras, but non-vanishing in relevant infinite-dimensional cases. The extension is described in a few different gradings, where it is given a covariant description with respect to different subalgebras. We expect the results to be important for the connection between extended geometry and cosmological billiards.
\end{quote} 

\vfill

\hrule

\noindent{\tiny email:
  martin.cederwall@chalmers.se, jakob.palmkvist@chalmers.se}

\newpage

\tableofcontents

\section{Introduction\label{IntroSec}}

Gravity (and models containing gravity) exhibit hidden symmetries on dimensional reduction to 3 or 2 dimensions,
which serve as solution-generating groups for solutions with the appropriate amount of Killing vectors.
These are the famous Ehlers \cite{Ehlers:1957zz} and Geroch \cite{Geroch:1972yt} symmetries.
The Ehlers symmetry for $D$-dimensional Einstein gravity reduced to 3 dimensions is $SL(D-2)$, which may be enhanced by the presence of non-gravitational fields; for $D=11$ supergravity it is $E_{8(8)}$. The Geroch symmetries obtained on further reduction to 2 dimensions are (untwisted) affine Kac--Moody (KM) extensions of the Ehlers symmetry, thus infinite-dimensional.

Further extension leads to an over-extended KM group, often hyperbolic. 
It does not simply occur on reduction to 1 dimension
as an extension of the Geroch symmetry, but is realised in a different way.
Close 
to a space-like singularity in the unreduced theory, time-derivatives dominate over spatial gradients, and spatial separation in the limit implies decoupling \cite{BKL}. In this
Belinskii--Khalatnikov--Lifshitz (BKL) limit,
gravity has been conjectured to be described by so called cosmological billiards, amounting to particle motion 
in a Weyl chamber of an over-extended KM algebra \cite{Damour:2001sa,Damour:2002mp,Damour:2002et,Henneaux:2007ej}.
An important and necessary observation making this possible is the presence of gradient structures in 
over-extended KM algebras \cite{Damour:2002cu,Kleinschmidt:2005bq}. 
A level expansion preserving a $\gl(d)$ subalgebra shows that the algebra is ``big enough'' to contain infinite sequences of modules possible to interpret as increasing numbers of spatial gradients.

The framework for making all these symmetries appearing in gravity (or models containing gravity, \eg\ supergravities) ``unhidden''
is extended geometry \cite{Cederwall:2017fjm,Cederwall:2018aab,Cederwall:2019qnw,Cederwall:2019bai,Cederwall:2021xqi,Cederwall:2023xbj,Bossard:2024gry}, a unified framework encompassing \eg\ double geometry
\cite{Tseytlin:1990va,Siegel:1993xq,Siegel:1993bj,Hitchin:2010qz,Hull:2004in,Hull:2006va,Hull:2009mi,Hohm:2010jy,Hohm:2010pp,Jeon:2012hp,Park:2013mpa,Berman:2014jba,Cederwall:2014kxa,Cederwall:2014opa,Cederwall:2016ukd}
and exceptional geometry \cite{Hull:2007zu,Pacheco:2008ps,Hillmann:2009pp,Berman:2010is,Berman:2011pe,Coimbra:2011ky,Coimbra:2012af,Berman:2012vc,Park:2013gaj,Cederwall:2013naa,Cederwall:2013oaa,Aldazabal:2013mya,Hohm:2013pua,Blair:2013gqa,Abzalov:2015ega,Hohm:2013vpa,Hohm:2013uia,Hohm:2014fxa,Cederwall:2015ica,Bossard:2017wxl,Bossard:2017aae,Bossard:2018utw,Bossard:2019ksx,Bossard:2021jix,Bossard:2021ebg,Bossard:2023ajq}.
This has been done for Ehlers \cite{Hohm:2013jma,Hohm:2014fxa} and Geroch 
\cite{Bossard:2017aae,Bossard:2018utw,Bossard:2021jix,Bossard:2024gry}
symmetries, but yet not completely for BKL symmetry (see however ref. \cite{Bossard:2021ebg} for a partial construction involving a very extended KM algebra).

A crucial ingredient in extended geometry is tensor hierarchy algebras \cite{Palmkvist:2013vya,Carbone:2018xqq,Cederwall:2019qnw,Cederwall:2021ymp}. 
Gradings of such superalgebras inform us on the content of fields, ghosts and antifields in the different models, and 
also provide a means of writing relevant brackets as derived brackets
\cite{Palmkvist:2013vya,Palmkvist:2015dea,Cederwall:2019bai}, leading to a Batalin--Vilkovisky formulation of the dynamics
\cite{Cederwall:2018aab,Cederwall:2019bai,Cederwall:2023xbj}.
In ref. \cite{Cederwall:2021ymp}, the complete structure of the tensor hierarchy algebra extension of an over-extended KM algebra was conjectured, with very strong support from the counting of modules in certain gradings in examples. As a vector space, it contains two shifted copies of the corresponding Borcherds superalgebra\footnote{For finite-dimensional $\fg$, the Borcherds superalgebra $\BB(\fg,\lambda)$ is a contragredient superalgebra containing $\fg\oplus\CC$ at degree 0 (in a grading with respect to the ``fermionic'' node), the modules $R(\mp\lambda)$ at degrees $\pm1$, and covariant Serre relations in $R(\mp2\lambda)$ at degrees $\pm2$. The tensor hierarchy algebra $S(\fg,\lambda)$ contains $\fg$ at degree 0, $R(-\lambda)$ at degree 1, and the same relations at degree 2. At degree $-1$, the maximal module respecting the degree 2 relations appears, the so called embedding tensor module. We refer to ref. \cite{Cederwall:2019qnw} for details. When $\fg$, is infinite-dimensional, new modules appear in the tensor hierarchy algebra \cite{Cederwall:2021ymp}.} 
$\BB(\fg^{++},\lambda)$: $S(\fg^{++},\lambda)=\BB(\fg^{++},\lambda)\oplus\BB(\fg^{++},\lambda)[1]$.
One surprising consequence is that at level 0, the over-extended KM algebra itself gets extended by generators in its fundamental module, and that there thus is a Lie algebra on this vector space.

The main purpose of this paper is to connect two of these phenomena: The ``extra generators'' appearing in the tensor hierarchy algebras, and the gradient structures. We will show how the gradient structures are generated by successively applying the ``extra'' transformations.
We expect this to become an important ingredient in establishing a connection between BKL extended geometry and billiards.

Physically relevant over-extended KM algebras tend to be hyperbolic. 
{\it A priori}, we do not see any particular r\^ole of hyperbolicity in our construction, apart from possibly in the tensorial decompositions of 
Section \ref{CovDecSec}.

In Section \ref{ExtensionSection}, we will describe the Lie algebra extension of an over-extended KM algebra by its fundamental module, including 
the identification of relevant Lie algebra cohomologies of the unextended and extended algebras. Section \ref{GradientSection} deals with the precise sense in which gradients are generated by the extra module, the covariance of the extension is discussed and examples of different gradings are given. We end by giving an outlook in Section \ref{OutlookSection}.



\begin{figure}
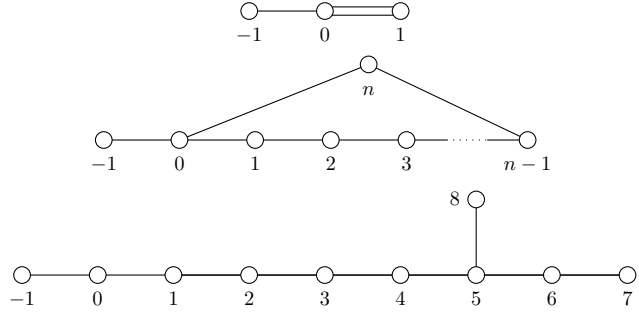

\begin{center}
\begin{dynkinDiagram}[arrows=false,labels={-1,0,1}]B{ooo}
\end{dynkinDiagram}\\
\begin{dynkinDiagram}[name=lower,labels={-1,0,1,2,3,n-1}]A{ooooo.o}
\node (current) at ($(lower root 4)+(0.5cm,1cm)$) {}; 
\dynkin[at=(current),name=upper,labels={n}]A{o}
\begin{pgfonlayer}{Dynkin behind}
	\draw[/Dynkin diagram] 
	($(upper root 1)$) -- ($(lower root 2)$);
\end{pgfonlayer}
\begin{pgfonlayer}{Dynkin behind}
	\draw[/Dynkin diagram] 
	($(upper root 1)$) -- ($(lower root 6)$);
\end{pgfonlayer}
\end{dynkinDiagram}\\
\begin{dynkinDiagram}[name=left,labels={-1,0}]A{oo}
\node (current) at ($(left root 2)+(7cm,0cm)$) {}; 
\dynkin[at=(current),name=right,backwards=true,labels={7,8,6,5,4,3,2,1}]E{oooooooo}
\begin{pgfonlayer}{Dynkin behind}
	\draw[/Dynkin diagram] 
	($(left root 2)$) -- ($(right root 1)$);
\end{pgfonlayer}
\end{dynkinDiagram}
\caption{Dynkin diagrams of $\fg^{++}$ for $\fg=A_1,A_n,E_8$. The Dynkin diagrams of $\fg^+$ are obtained by deleting node $-1$ and those of $\fg$ by also deleting node $0$.}
\end{center}
\label{DynkinFigure}
\end{figure}

Some conventions used throughout the paper are: 
\begin{itemize}
\item The ground field can be taken as $\CC$ (although physical applications use $\RR$, typically with split real form of the occurring KM algebras).
\item $\fg$ denotes a finite-dimensional simple Lie algebra, $\fg^+$ its untwisted affine extension and $\fg^{++}$ the subsequent over-extension. 
\item We denote highest weight modules $R(\lambda)$ and lowest weight ones $R(-\lambda)$, for $\lambda$ an integral dominant weight. Fundamental weights dual to simple roots $\alpha_i$ are $\Lambda_i$
labelled by the index $i$ according to Fig.~\ref{DynkinFigure}.
\item The semi\-direct sum of a Lie algebra $\fa$ and its module $R$ is denoted $\fa\inplus R$. Symmetric and antisymmetric parts of tensor products are denoted $\vee$ and $\wedge$, respectively.
\item Level $i$ of a $\ZZ$-graded vector space $V$ is denoted $V_i$. Shifts are defined so that $V[i]_j=V_{i+j}$.
\end{itemize}



\section{Extensions of hyperbolic algebras\label{ExtensionSection}}

\subsection{Virasoro extensions of affine Kac--Moody algebras}

As a preamble, let us revisit derivations and extensions of affine KM algebras.
Unlike finite-dimensional (semi-)simple Lie algebras, they have outer derivations, spanning a Virasoro algebra.
Some facts concerning affine KM algebras are collected in Appendix  \ref{AffineAppendix}.	
This is the ``first'' example of infinite-dimensional KM algebras having non-trivial Lie algebra cohomologies enabling such extensions, in this case $H^1(\fg^+;\fg^+)$. 

For the purpose of extending over-extended KM algebras, we are interested in the extension of affine KM algebras\footnote{We adhere to a convention where $L_0$ is not included in the algebra $\fg^+$, see Appendix \ref{AffineAppendix}.} by 
the subalgebra of the Virasoro algebra generated by $L_0$ and $L_1$. We rename the former to $\dd$, since its action can be shifted.
Specifying a module of the extended algebra $\langle\dd\rangle\lplus\fg^{+}$ amounts to specifying a module of $\fg^+$ together with a number (scaling weight)
$s$ such that $\dd$ acts as $L_0+s$.
Note that although the extension of $\langle\dd\rangle\lplus\fg^+$ by $\langle L_1\rangle$ is a semidirect sum
$\langle\dd,L_1\rangle\lplus\fg^+$, it is not the extension by a (scalar) derivation of $\langle\dd\rangle\lplus\fg^+$, since $[\dd,L_1]=-L_1$. Rather it is a ``transforming derivation'' by a scalar density. 
This statement can be formalised to stating that, with $\fa=\langle\dd\rangle\lplus\fg^+$, the relevant cohomology for the extension is not
$H^1(\fa;\fa)$, but $H^1(\fa;\fa\otimes\CC[1])=H^1(\fa;\fa[1])$, where the shift is in mode number (minus eigenvalue of $\dd$).
(There is also cohomology in $H^1(\fa;\fa[m])$, $m\in\ZZ$, $m\neq0$, corresponding to extension by $L_m$.)

This observation is important for the extension of an over-extended KM algebra, since $\dd$ becomes a Cartan generator in $\fg^{++}$ and $L_1$ the lowest state of a fundamental $F=R(-\Lambda_{-1})$.


\begin{figure}
\begin{center}
\includegraphics[scale=1.2]{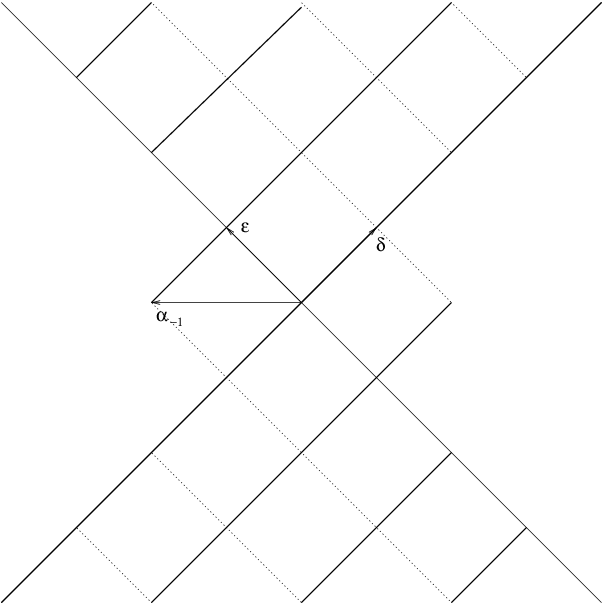}
\caption{\it The branching of an over-extended algebra $\fg^{++}$ into
modules of two affine $\fg^+$ algebras, indicated by the solid and dotted
diagonal lines. Each 
point in the diagram contains a finite-dimensional, completely reducible
$\fg$-module. The roots $\delta$ and $\epsilon$ are light-like, and $(\delta,\epsilon)=-1$.
The extending simple roots are $\alpha_0=\delta-\theta$ and $\alpha_{-1}=\epsilon-\delta$,
where $\theta$ is the highest root of $\fg$.
\label{HyperbolicFigure}}    
\end{center}
\end{figure}

\begin{figure}
\begin{center}
\includegraphics[scale=1.2]{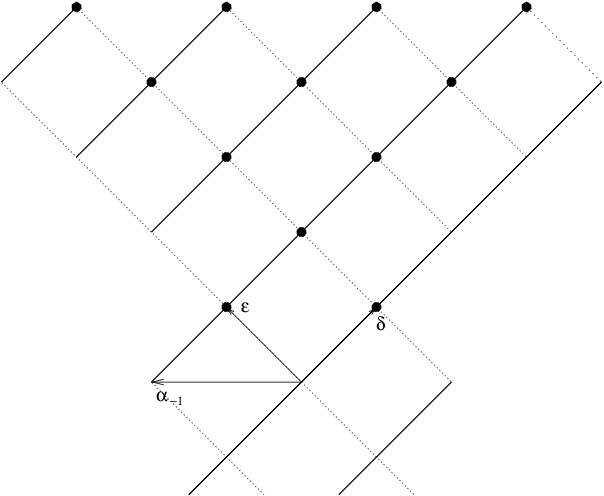}
\caption{\it The projection of weights (black dots) in the over-extended
  lowest weight fundamental module to the $\delta\epsilon$-plane. Note that $-\Lambda_{-1}=\delta$.}    
\label{HyperbolicFundFigure}
\end{center}
\end{figure}

\subsection{Extending over-extended KM algebras by the fundamental\label{FundExtSec}}

Affine subalgebras are of course subalgebras of over-extended KM algebras, but there is no canonical embedding.
Each choice also defines a Virasoro algebra, in particular a translation $L_1$. As $L_1$ is the lowest weight state in the $\fg^{++}$ module
$R(-\Lambda_{-1})$, any other light-like weight in $R(-\Lambda_{-1})$, related to $-\Lambda_{-1}$ by a transformation in the Weyl group, is an equally good translation. It seems natural to consider an extension by generators in the full $F=R(-\Lambda_{-1})$.

Extensions of over-extended KM algebras by its fundamental module were considered in ref. \cite{Cederwall:2021ymp}. It was observed that level 0 of a tensor hierarchy algebra extension of an over-extended KM algebra $\fg^{++}$ consists of a Lie algebra on the vector space $\fg^{++}\oplus F$, where $F$ generates a non-trivial action on $\fg^{++}$. 

The property that an element in $F$ {\it both} transforms under $\fg^{++}$ as a representation module {\it and} 
generates a transformation on $\fg^{++}$, implies that, for basis elements $T_\alpha\in\fg^{++}$ and $J_M\in F$, one needs a Lie bracket
\begin{align}
[T_\alpha,J_M]=-t_{\alpha M}{}^NJ_N+u_{M\alpha}{}^\beta T_\beta\;.
\end{align}
As remarked above, this is a necessary structure, reflected already in the subalgebra at $k=0$.

Let us formulate the existence of such ``transforming derivations'' in terms of Lie algebra cohomology.

A collection of ``derivations'' in the fundamental $F=R(-\lambda)$ acting on $\fg^{++}$ is a linear map
with coefficients $u_{M\alpha}{}^\beta$ in $\Hom(\fg^{++}\otimes F,\fg^{++})=\Hom(\fg^{++},\fg^{++}\otimes\bar F)$.
Non-trivial (outer) such maps should be characterised by $H^1(\fg^{++};\fg^{++}\otimes\bar F)$ (note that outer derivations of a Lie algebra $\fa$ are classified by 
$H^1(\fa;\fa)$).
Let the coalgebra basis elements (1-forms) be $c^\alpha$. The differential on elements in the complex\footnote{For a Lie algebra $\fg$ with generators $T_\alpha$, $C^\bullet(\fg)$ denotes the complex (freely) generated by the coalgebra 1-forms $c^\alpha$.},
$\omega\in \fg^{++}\otimes\bar F\otimes C^\bullet(\fg^{++})$, is
\begin{align}
(d\omega)_M{}^\alpha=-t_{\beta M}{}^Nc^\beta\omega_N{}^\alpha
-f_{\beta\gamma}{}^\alpha c^\beta\omega_M{}^\gamma
+\tfrac12f_{\beta\gamma}{}^\delta c^\beta c^\gamma{\*\over\*c^\delta}\omega_M{}^\alpha\;.
\end{align}
In index-free notation, this is 
\begin{align}
d=-\Delta(c)+\tfrac12\imath_{[c,c]}\;,
\end{align}
where $\Delta(a)v$ for any element $a$ in the Lie algebra and $v$ in a module is the action of $a$ in the module, \ie, 
$\Delta(a)v_I=a^\alpha t_{\alpha I}{}^J v_J$. When $a=T_\alpha$, we simply write $\Delta_\alpha=\Delta(T_\alpha)$,
so that \ie, $\Delta_\alpha v_I=t_{\alpha I}{}^J v_J$.
We see that an exact 1-form is 
\begin{align}
d\lambda_M{}^\alpha=-\Delta(c)\lambda_M{}^\alpha=-c^\beta(t_{\beta M}{}^N\lambda_N{}^\alpha+f_{\beta\gamma}{}^\alpha\lambda_M{}^\gamma)\,.
\end{align}
Thinking of the map as generated by $J_M$, the action on $\fg^{++}$ can be undone by a redefinition 
$J_M\rightarrow J_M+\lambda_M{}^\alpha T_\alpha$.
In this framework, we will have a Lie bracket in the extended algebra that includes
\begin{align}
[T_\alpha,T_\beta]&=f_{\alpha\beta}{}^\gamma T_\gamma\;,\nn\\
[T_\alpha,J_M]&=-t_{\alpha M}{}^NJ_N+u_{M\alpha}{}^\beta T_\beta\;.\label{TTJTbrackets}
\end{align}
A closed 1-form $\omega_M{}^\alpha=c^\beta u_{M\beta}{}^\alpha$ obeys $(d\omega)_M{}^\alpha=0$, which spelled out
reads
\begin{align}
f_{\beta\gamma}{}^\delta u_{M\delta}{}^\alpha+2u_{N[\beta}{}^\alpha t_{\gamma]M}{}^N
+2u_{M[\beta}{}^\delta f_{\gamma]\delta}{}^\alpha=0\;.
\end{align} 
This is exactly the $T$ part of the Jacobi identity $[T,T,J]$, using eq. \eqref{TTJTbrackets}.

 Finding a $u\in H^1(\fg^{++};\fg^{++}\otimes\bar F)$ is a linear problem. 
 It can also be understood as (part of) a deformation problem for the Lie algebra $\fg^{++}\lplus F$, the solutions of which are classified
 by $H^2(\fg^{++}\lplus F;\fg^{++}\lplus F)$. Under the assumption that a cocycle can be found, such that the bracket of two $J$'s is deformed to
\begin{align}
[J_M,J_N]=g_{MN}{}^PJ_P \label{JJJbrackets}
\end{align}
 (\ie, the ``derivations'' form a Lie subalgebra, which was argued in ref. \cite{Cederwall:2021ymp}), this cohomology also implies the identity
 \begin{align}
 g_{MN}{}^Q t_{\alpha Q}{}^P
    +2t_{\alpha[M}{}^Q g_{N]Q}{}^P
    +2t_{\beta[M}{}^P u_{N]\alpha}{}^\beta=0\;,\label{gtturel}
 \end{align}
 which is the $[T,J,J]$ Jacobi identity.
A full non-linear deformation, leading to a Lie algebra $\mathscr{A}$ on the vector space $\fg^{++}\oplus F$,
with brackets (\ref{TTJTbrackets}) and (\ref{JJJbrackets}),
also demands
$[u_M,u_N]_\alpha{}^\beta=g_{MN}{}^Pu_{P\alpha}{}^\beta$ and the Jacobi identities for $g_{MN}{}^P$.
We thus have the complete set of Jacobi identities for $\mathscr{A}$:
\begin{align}
0&=  f_{[\alpha\beta}{}^\epsilon
  f_{\gamma]\epsilon}{}^\delta\;,\nn\\
0&=  [t_\alpha,t_\beta]_M{}^N-f_{\alpha\beta}{}^\gamma t_{\gamma
  M}{}^N\;,\nn\\
0&=  f_{\alpha\beta}{}^\delta u_{M\delta}{}^\gamma
    +2u_{M[\alpha}{}^\delta f_{\beta]\delta}{}^\gamma
    +2u_{N[\alpha}{}^\gamma t_{\beta]M}{}^N\;,\nn\\
0&=  g_{MN}{}^Q t_{\alpha Q}{}^P
    +2t_{\alpha[M}{}^Q g_{N]Q}{}^P
    +2t_{\beta[M}{}^P u_{N]\alpha}{}^\beta\;,\\
0&=  [u_M,u_N]_\alpha{}^\beta-g_{MN}{}^Pu_{P\alpha}{}^\beta
    \;,\nn\\
0&=  g_{[MN}{}^Rg_{P]R}{}^Q\;.\nn
\end{align}

It should be mentioned that a strict proof, in particular of the statement that $F$ can be chosen to form a subalgebra, is lacking. 
We refer to 
Appendix \ref{ConjProofAppendix} for more comments on proofs vs. conjectures.

\subsection{Representations and cocycles}

We will sometimes use a (lower) index $\hat\alpha$ for basis elements of $\mathscr{A}$,
and ${\hat{\Delta}}_{\hat\alpha}$ for the adjoint action of these basis elements in $\mathscr{A}$.

It was observed in ref. \cite{Cederwall:2021ymp} that $\mathscr{A}$ is linearly represented on $F=R(-\lambda)$.
(This follows from the occurrence of $R(\pm\lambda)$ at certain levels in a tensor hierarchy algebra.)
The corresponding representation matrices are $t_{\alpha M}{}^N$ and $j_{MN}{}^P$, fulfilling
\begin{align}
[t_\alpha,t_\beta]&=f_{\alpha\beta}{}^\gamma t_\gamma\;,\nn\\
[t_\alpha,j_M]&=-t_{\alpha M}{}^Nj_N+u_{M\alpha}{}^\beta t_\beta\;,\label{fundtrels}\\
[j_M,j_N]&=g_{MN}{}^Pj_P\;.\nn
\end{align}
The fundamental now appears in two distinct ways, as a subspace (a subalgebra) of $\mA$ and as a module over $\mA$, with different transformation properties.
To make the distinction, when necessary, we write the former index dotted, so $\hat\alpha=(\alpha,\dot M)$.

It is peculiar that there are two objects with the same index structure, $g_{MN}{}^P$ and $j_{[MN]}{}^P$. They turn out to be related.
Consider the cocycle identity \eqref{gtturel}. It can be written in terms of the $\fg^{++}$ transformation of $g$ as
$\Delta_\alpha g_{MN}{}^P=2t_{\beta[M}{}^Pu_{N]\alpha}{}^\beta$.
On the other hand, the middle equation in \eqref{fundtrels} is expressed as
$\Delta_\alpha j_{MN}{}^P=t_{\beta N}{}^Pu_{M\alpha}{}^\beta$.
Thus, $g_{MN}{}^P+2j_{[MN]}{}^P$ is a $\fg^{++}$ tensor, $\Delta_\alpha(g_{MN}{}^P+2j_{[MN]}{}^P)=0$.
Since $\wedge^2R(-\lambda)\not\supset R(-\lambda)$, it vanishes, and 
\begin{align}
g_{MN}{}^P=-2j_{[MN]}{}^P\;.
\end{align}

This relation implies the existence of a certain cocycle in $H^1(\mA,F)$, namely the projection on the fundamental part of $\mA$.
Let us denote the structure constants of $\mA$ ($f$, $t$, $u$ and $g$) collectively as $\hat f_{\hat\alpha\hat\beta}{}^{\hat\gamma}$ and the representation matrices on $F$ ($t$ and $j$) as $\hat t_{\hat\alpha M}{}^N$.
The differential on $F\otimes C^\bullet(\mA)$ is
$d=-\hat\Delta(c)+{1\over2}\langle[c,c],{\*\over\*c}\rangle$. Acting on a 1-form $\omega^M=\delta_{\hat\alpha}^Mc^{\hat\alpha}=c^M$:
\begin{align}
(d\omega)^M&=c^{\hat\alpha}\hat t_{\hat\alpha P}{}^Mc^P+{1\over2}c^{\hat\alpha}c^{\hat\beta}\hat f_{\hat\alpha\hat\beta}{}^M\nn\\
&=(t_{\alpha P}{}^Mc^\alpha c^P+j_{NP}{}^Mc^Nc^P)+(-t_{\alpha P}{}^Mc^\alpha c^P+\tfrac12g_{NP}{}^Mc^Nc^P)\\
&=\tfrac12(g_{NP}{}^M+2j_{[NP]}{}^M)c^Nc^P=0\;.\nn
\end{align}
$\omega^M$ is obviously not exact, so it represents non-trivial cohomology.
Note that the cocycle projects on the fundamental part of $\mA$ and then ``reinterprets'' the fundamental as an element in the fundamental module of $\mA$, thus converting an index $\dot M$ to $M$. 
This cocycle will be used in the following Section for the concrete construction of gradients.

\section{Gradients from extensions\label{GradientSection}}

\subsection{Back to \texorpdfstring{$\fg^{++}\lplus F$}{}}

The cocycle in $H^1(\mA,F)$ of the previous Section can be used to build indecomposable representations of $\mA$ (some of which in fact appear in the tensor hierarchy algebra). 

The construction we want to use is the following.
Consider the transformations under $T$ and $J$ of an element $X$ in some module $R$,
represented by $\hat\Delta_\alpha X$ and $\hat\Delta_M X$.
Then after acting with the latter, the $M$ index is reinterpreted as a fundamental index, so $\hat\Delta_M X$ is seen as an element in $\bar F\otimes R$.
Then, since $[\hat\Delta_{\hat\alpha},\hat\Delta_{\hat\beta}]=-f_{\hat\alpha\hat\beta}{}^{\hat\gamma}\Delta_{\hat\gamma}$,
\begin{align}
[\hat\Delta_\alpha,\hat\Delta_\beta]&=-f_{\alpha\beta}{}^\gamma\hat\Delta_\gamma\;,\nn\\
[\hat\Delta_\alpha,\hat\Delta_M]&=[\hat\Delta_\alpha,\hat\Delta_{\dot M}]+u_{M\alpha}{}^\beta\hat\Delta_\beta
=-f_{\alpha M}{}^{\hat\beta}\hat\Delta_{\hat\beta}+u_{M\alpha}{}^\beta\hat\Delta_\beta=t_{\alpha M}{}^N\hat\Delta_N\;, \label{DeltaCommutators}\\
[\hat\Delta_M,\hat\Delta_N]&=[\hat\Delta_{\dot M},\hat\Delta_{\dot N}]+2(g_{MN}{}^P+j_{[MN]}{}^P)\hat\Delta_P
=(g_{MN}{}^P+2j_{[MN]}{}^P)\hat\Delta_P=0\;,\nn
\end{align}
which is the algebra $\fg^{++}\lplus F$.
The ``flatness'' property expressed by the last equation is crucial for its interpretation in terms of translations (gradient structures).
Repeated transformations $\hat\Delta_M$ gives an arbitrary number of (automatically symmetric) indices,
$X_{M_1\cdots M_k}=\hat\Delta_{M_1}\cdots\hat\Delta_{M_k}X$.

\subsection{Gradients}

The semidirect sum $\fg^{++}\lplus F$ is recovered in eq. \eqref{DeltaCommutators}, and the commutative action of $\hat\Delta_M$ appends fundamental indices on any object. 
It should be stressed that although the algebra of the $\hat\Delta_M$'s is abelian, they generate non-trivial transformations.
It is now tempting to try to interpret $\hat\Delta_M$ as a ``derivative'', providing a gradient structure. 

In extended geometry, derivatives are declared to lie in a module $\bar F=R(\lambda)$ of some structure group $\underline{G}$ with Lie algebra 
$\underline{\fg}$.
(We are here particularly interested in the case $\underline{\fg}=\fg^{++}$, but $\underline{\fg}$ may be any
Kac--Moody algebra, finite- or infinite-dimensional, not necessarily over-extended.)
They are then subject to a so called (strong) section constraint that implies that demands that products of derivatives (momenta) only contain the leading symmetric and antisymmetric modules in $\otimes^2R(\lambda)$. Concretely \cite{Bossard:2017aae,Cederwall:2017fjm},
\begin{align}
(-\eta^{\alpha\beta}t_\alpha\otimes t_\beta+(\lambda,\lambda)-1+\sigma)\*\otimes\*=0\;,
\end{align}
where $\sigma$ is a permutation operator that interchanges the two derivatives.
Solutions to the section constraints are maximal linear subspaces of the minimal $\underline{G}$-orbit of $R(\lambda)$ (``sections''). 
When $\lambda$ is a fundamental weight,
they can be found graphically by following a ``gravity line'' in the Dynkin diagram, starting at the node corresponding
to $\lambda$.  Consequently, a choice of section, corresponding to an anchor map in generalised geometry, breaks $\underline{G}$ to a $GL(n)$ subgroup, corresponding to ordinary (``non-extended'') physical momenta.
The {\it weak} section constraint is the symmetric part, obtained when derivatives act on the same object, and components of momentum simply are multiplied.
Its solutions give the minimal orbit. The commutative ring of momenta obeying the weak section constraint---the coordinate ring of the minimal orbit---is Koszul dual to the positive levels of a Borcherds superalgebra, $\BB(\underline{\fg},\lambda)$ (closely related to the tensor hierarchy algebra 
$W(\underline{\fg},\lambda)$ and coinciding with it at positive levels when $\underline{\fg}$ is finite-dimensional), see \eg\ refs. \cite{Cederwall:2015oua,Cederwall:2023wxc}.

A question that now arises is: Is the ``derivative'' $\hat\Delta_M$, acting on some module, unconstrained, or does it in some sense satisfy a constraint? The strongest possible such constraint, and also the most interesting, since it relates to extended geometry, would be the weak section constraint.
Acting in $\fg^{++}$, the weak section constraint would amount to the identity
\begin{align}
\eta^{\gamma\delta}t_{\gamma M}{}^P t_{\delta N}{}^Q(u_{(P}u_{Q)}+j_{(PQ)}{}^Ru_R)_\alpha{}^\beta\overset{?}=0\;.
\label{SectionQuestionEq}
\end{align}
We have not been able to find general identities (among those coming from Jacobi identities) that would help to check this.
An alternative strategy would be to perform explicit checks in some grading.
The simplest one to use is the affine grading of Section~\ref{AffineGradingSection} and Appendix~\ref{AffineDecompSec}, because there the subleading symmetric module starts appearing already at level 2, while in the other gradings it appears at higher levels.
We have tried to check the weak section constraint by inserting the subleading combination of two $\hat\Delta_M$'s appearing in the first equation of
\eqref{LeadingSubleading}, but have so far obtained non-zero results.

If this conclusion holds, it thus seems that $\hat\Delta_M$ contains more than gradients corresponding to some section,
but exact statements, given the module they act on, are lacking (if it is not simply the whole module), and the precise interpretation remains unclear.
However, commutativity of course allows restricting to $\hat\Delta_m$ in some section to obtain gradient structures.

\subsection{Covariance and decompositions of $\mA$\label{CovDecSec}}

Cocycles are not tensors. There is no way to write the Lie algebra $\mA$ in terms of $\fg^{++}$-invariant tensors (structure constants and representation matrices). It is still of course desirable to have some concrete expressions. What, then, is the maximal manifest symmetry?
We have seen that the cocycle first arises for affine KM algebras, and that the extended algebra then can be given a formulation where the affine symmetry is manifest, due to the presence of Virasoro generators. (The situation there is really not different, however the action of Virasoro generators, though formally ``non-tensorial'', is under control.)
The Lie algebra $\mA$ can be constructed level by level in a grading with respect to the over-extending node, manifesting affine covariance. 

There are many other gradings. Take a grading with respect to another node, any node if $\fg^{++}$ is hyperbolic, otherwise a node such that its removal yields a Dynkin diagram for a finite-dimensional or affine KM algebra. That subalgebra will then not exhibit the Lie algebra cohomology representing extensions of the kind we are dealing with, and can be kept manifest in a level by level construction of $\mA$.
Different choices of grading, \ie, different manifest subalgebras, will correspond to different representatives in the Lie algebra cohomology in question, meaning that they differ by redefinitions of the generators $J_M$.  
These redefinitions should be such that they preserve the property $[F,F]\subset F$. We have not tried to show this in general, but note that in examples this condition fixes the brackets when there are common modules in $\fg^{++}$ and $R(-\Lambda_{-1})$.

In the following subsections, $\mA$ and its fundamental representation are constructed for a few levels in some examples.
We also illustrate the first instances of redefinitions of generators (change of representative in cohomology) when switching from one grading node to a neighbouring one.

\subsubsection{$\fg^+\oplus\CC$ decomposition\label{AffineGradingSection}}

In order to investigate the decomposition of $\mA$ in modules of a subalgebra $\fg^+\oplus\CC\subset\fg^{++}$,
we use the decomposition of $\fg^{++}$ and $R(-\Lambda_{-1})$ of Appendix \ref{AffineDecompSec}.
As described there, degree $1$ of $\fg^{++}$ contains a shifted affine fundamental module, with basis elements $T_\mu$.
The lowest weight fundamental of $\fg^{++}$ contains the lowest weight state $J$ at degree 0, and an unshifted fundamental at degree 1 with basis elements $J_\mu$.
Starting from the lowest basis vector $J$ and its identification with $L_1$, we immediately have
\begin{align}
[J,\bar T^\mu]&=(\ell_1)_\nu{}^\mu\bar T^\nu\;,\nn\\
[J,T_\mu]&=-(\ell_1)_\mu{}^\nu T_\nu-J_\mu\;.
\end{align}
Using these, a short calculation using the Jacobi identity $[J,T_\mu,\bar T^\nu]$ yields
\begin{align}
[J_\mu,\bar T^\nu]=-\delta_\mu^\nu J-(\ell_1)_\mu{}^\nu\KK+\sum_{m\in\ZZ}\eta^{ab}(t_{a,m})_\mu{}^\nu T_{b,1-m}\;.
\end{align}
The Jacobi identity $[J,J_\mu,\bar T^\nu]$ is then seen to necessitate
\begin{align}
[J,J_\mu]=-(\ell_1)_\mu{}^\nu J_\nu\;.
\end{align} 
The level one elements span an indecomposable module.

It is straightforward to continue to level 2. The result is, as expected, unique:
\begin{align}
[T_\mu,T_\nu]&=2T_{\mu\nu}\;,\nn\\
[T_\mu,J_\nu]&=-(C_1(C_0-2)^{-1})_{\mu\nu}{}^{\kappa\lambda}T_{\kappa\lambda}+J_{\mu\nu}\;,\\
[J_\mu,J_\nu]&=-2(C_1(C_0-2)^{-1})_{[\mu\nu]}{}^{\kappa\lambda}J_{\kappa\lambda}\;,\nn
\end{align}
together with
\begin{align}
[T_{\mu\nu},\bar T^\kappa]&=(C_0-2)_{[\mu\nu]}{}^{\kappa\lambda}T_\lambda\;,\nn\\
[J_{\mu\nu},\bar T^\kappa]&=(C_1)_{(\mu\nu)}{}^{\kappa\lambda}T_\lambda-(C_0-1+\sigma)_{\mu\nu}{}^{\kappa\lambda}J_\lambda\;.
\end{align}
The inverses are well defined on the object they act on (the kernels are in the ideals removed in Appendix \ref{AffineDecompSec}).
All Jacobi identities resulting in level 2 or lower are checked.
This implies the action of $J$ on level 2:
\begin{align}
[J,T_{\mu\nu}]&=-((C_0-2)(\ell_1\otimes1+1\otimes\ell_1)(C_0-2)^{-1})_{\mu\nu}{}^{\kappa\lambda}T_{\kappa\lambda}-J_{[\mu\nu]}\;,\nn\\
[J,J_{\mu\nu}]&=-(\ell_1\otimes1+1\otimes\ell_1)_{(\mu\nu)}{}^{\kappa\lambda}J_{\kappa\lambda}\\
	&\qquad-((C_0-2)(\ell_1\otimes1+1\otimes\ell_1)(C_0-2)^{-1})_{[\mu\nu]}{}^{\kappa\lambda}J_{\kappa\lambda}\;.\nn
\end {align}
Note that the action is $\ell_1\otimes1+1\otimes\ell_1$ in some basis, along with the ``shift term'' (last term in first equation).

We can also find the low levels of the action of $\mA$ in its fundamental representation.
Along with the representation matrices of the generators in $\fg^{++}$ given in Appendix \ref{AffineDecompSec},
\begin{align}
J\cdot E&=0\;,\nn\\
J\cdot E_\mu&=-(\ell_1)_\mu{}^\nu E_\nu\;,\nn\\
J_\mu\cdot E&=0\;,\\
J\cdot E_{\mu\nu}&=-((C_0-1+\sigma)(\ell_1\otimes1+1\otimes\ell_1)(C_0-1+\sigma)^{-1})_{\mu\nu}{}^{\kappa\lambda}E_{\kappa\lambda}\;,\nn\\
J_\mu\cdot J_\nu&=-(C_1(C_0-1+\sigma)^{-1})_{\mu\nu}{}^{\kappa\lambda}E_{\kappa\lambda}\;,\nn\\
J_{\mu\nu}\cdot E&=(C_1C_0^{-1})_{(\mu\nu)}{}^{\kappa\lambda}E_{\kappa\lambda}\;.\nn
\end{align}
Note that the relation $g_{MN}{}^P=-2j_{[MN]}{}^P$ is satisfied.

\subsubsection{$\gl(2)\oplus\fg$ decomposition}

This is the grading vertically in Figs. \ref{HyperbolicFigure} and \ref{HyperbolicFundFigure}.

Let us introduce the notation for $\fg$-modules: $\adj=R(\theta)$, $\vee^2\adj=R(2\theta)\oplus\sigma_2$.
Modules over $\sl(2)$ are named by dimension.
The first few levels of $\fg^{++}$ are
\begin{align}
(\fg^{++})_0&=(1,\adj)\oplus(3,1)\oplus(1,1)\;,\nn\\
(\fg^{++})_1&=(2,\adj)\;,\\
(\fg^{++})_2&=(3,\adj)\oplus(1,\sigma_2)\;,\nn\\
(\fg^{++})_3&=(4,\adj)\oplus(2,\wedge^2\adj\oplus\sigma_2)\;.\nn
\end{align}
The highest and lowest $\sl(2)$ states in $(m,\adj)\subset(\fg^{++})_m$ are loop generators $T_{a,m}$ in the two affine subalgebras indicated in Fig. \ref{HyperbolicFigure}.
The lowest weight $\fg^{++}$ fundamental starts as
\begin{align}
F_1&=(2,1)\;,\nn\\
F_2&=(1,\adj)\;,\\
F_3&=(2,\adj)\oplus(2,\sigma_2)\;.\nn
\end{align}
We will only give explicit expressions up to degree 2.

We already note that the first overlap between $\fg^{++}$ and $F$ seen as modules over the algebra $\fg$, which is the common subalgebra of an affine subalgebra and $\sl(2)\oplus\fg$, occurs as an adjoint at bidegree $(2,0)$, where the first degree is with respect to node 0 and the second the $\sl(2)$ weight. We therefore expect that a change of representative for the extension switching between affine covariance and $\sl(2)\oplus\fg$ covariance will involve a redefinition $J\rightarrow J+T$ in this module (as well as similar redefinitions in overlapping modules at higher degrees).

Obviously, all $(\fg^{++})_m$ are finitely reducible modules of $(\fg^{++})_0$. Since (unlike the affine decomposition) $F$ has no part in degree 0, there will be no indecomposable structures with respect to the degree 0 subalgebra.

Normalising the action of $(\fg^{++})_0=\langle T_a,T_i{}^j,H\rangle$ on $(\fg^{++})_1=\langle T^{(1)}_{ia}\rangle$ as
\begin{align}
[T_a,T^{(1)}_{ib}]&=f_{ab}{}^cT^{(1)}_{ic}\;,\nn\\
[T_i{}^j,T^{(1)}_{ka}]&=\delta^j_kT^{(1)}_{ia}-\tfrac12\delta^j_iT^{(1)}_{ka}\;,\\
[H,T^{(1)}_{ia}]&=T^{(1)}_{ia}\;,\nn
\end{align}
we have (modulo an overall normalisation)
\begin{align}
[T^{(1)}_{ia},T^{(-1)j}{}_ b]=\delta_i^jf_{ab}{}^cT_c+\eta_{ab}T_i{}^j-\tfrac12\eta_{ab}\delta_i^jH\;.
\end{align}
Then the Jacobi identities $[T^{(1)}, T^{(1)},T^{(-1)}]$ are consistent with the degree 2 ideals,
\begin{align}
[T^{(1)}_{ia},T^{(1)}_{jb}]=f_{ab}{}^cT^{(2)}_{ijc}+\epsilon_{ij}T^{(2)}_{ab}\;,
\end{align}
and
\begin{align}
[T^{(2)}_{ija},T^{(-1)k}{}_ b]&=-f_{ab}{}^c\delta_{(i}^kT^{(1)}_{j)c}\;,\nn\\
[T^{(2)}_{ab},T^{(-1)ic}]&=\epsilon^{ij}(2\delta_{(a}^c\delta_{b)}^d-f_{(a}{}^{ce}f_{b)}{}^d{}_e)T^{(1)}_{jd}
\end{align}
(the matrix in the the last equation annihilates the leading symmetric $R(2\theta)$).

In the fundamental representation, we have basis elements $E^{(1)}_i$ at degree 1 and $E^{(2)}_a$ at degree 2, and
\begin{align}
T^{(1)}_{ia}\cdot E^{(1)}_j=\epsilon_{ij}E^{(2)}_a\;,
\end{align}
which is consistent with
\begin{align}
T^{(-1)i}{}_a\cdot E^{(2)}_b=\eta_{ab}\epsilon^{ij}E^{(1)}_j\;.
\end{align}

We can now begin to construct $\mA$ in this grading, by declaring
\begin{align}
[T^{(-1)i}{}_a,J^{(1)}_j]=\delta_j^iT_a\:.
\end{align} 
The Jacobi identity $[T^{(-1)},T^{(1)},J^{(1)}]$ then determines uniquely
\begin{align}
[T^{(1)}_{ia},J^{(1)}_j]&=-T^{(2)}_{ija}+\epsilon_{ij}J^{(2)}_a\;,\label{T1J1Eq}\\
[T^{(-1)i}{}_ a,J^{(2)}_b]&=\tfrac12\epsilon^{ij}f_{ab}{}^cT^{(1)}_{jc}+\epsilon^{ij}\eta_{ab}J^{(1)}_j\;.
\end{align}

We would like to compare to the affine decomposition. Part of eq. \eqref{T1J1Eq} reads
$[T^{(1)}_{1a},J^{(1)}_2]=-T^{(2)}_{12a}+J^{(2)}_a$. $T^{(1)}_{1a}$ is identified with the loop generator $T_{a,1}$, while 
$J^{(1)}_2$ corresponds to the lowest state $|0\rangle_J$ in $J_\mu$. In the affine decomposition,
$[T_{a,1},J_\mu]=-(t_{a,1})_\mu{}^\nu J_\nu$. 
Let us denote the $J$'s in the $\sl(2)\oplus\fg$ decomposition with a prime.
The identification is $-T^{(2)}_{12a}+J'^{(2)}_a=J^{(2)}_a(=-t_{a,1}|0\rangle_J)$. 
Such redefinitions will certainly continue to appear at higher levels, where $T$ and $J$ contain overlapping $\fg$ modules,
and will be the redefinitions (change of the cocycle by an exact term) that relate decompositions with different covariance, defined by gradings with respect to different nodes.

\subsubsection{$\gl(d)$ decomposition\label{gldSubSection}}

This grading is particularly interesting, since the highest level in $\bar F$ provides a solution to the strong section constraint. The gradient structures are then generated by level 1.
Let us for simplicity consider $\gl(3)\subset A_1^{++}$, the grading is with respect to the rightmost node
in the Dynkin diagram at the top of Fig.~\ref{DynkinFigure}.
The end of this Section contains some comments on other cases.

\begin{table}
\begin{center}
{\small
\begin{tabular}{c|c|c||c|c|}
level&\multicolumn{2}{c||}{$A_1^{++}$}&\multicolumn{2}{c|}{$R(-\Lambda_{-1})$}\\ \hline
0&$(11)\oplus(00)$&$T_m{}^n,N$&&\\
1&$(02)$&$\Tn1^{mn}$&$(10)$&$\Jn1_m$\\
2&$(12)$&$\Tn2_m{}^{np}$&$(01)$&$\Jn2^m$\\
3&$(11)\oplus(22)$&$\Tn3_m{}^n,\Tn3_{mn}{}^{pq}$&$(03)\oplus(11)$&$\Jn3^{mnp},\Jn3_m{}^n$\\
4&$(02)\oplus(10)\oplus(13)\oplus2(21)\oplus(32)$&&$(02)\oplus(10)\oplus(13)\oplus2(21)$
\end{tabular}
}
\caption{Grading of $A_1^{++}$ and its lowest weight fundamental in terms of $\sl(3)$ modules, given by Dynkin labels, and notation for generators used in Section \ref{gldSubSection}.\label{A2levels}}
\end{center}
\end{table}

The content of the adjoint and of $R(-\Lambda_{-1})$ for a few levels are given in Table \ref{A2levels}.
We will give explicit expressions for brackets up to level 3.
The brackets with level 0 are given by
$[T_m{}^n,V_p]=-\delta^n_pV_m+\tfrac13\delta_m^nV_p$ and tensor product, and $[N,\Tn n]=n\Tn n$. This gives
\begin{align}
[\Tn1^{mn},\Tn{-1}_{pq}]&=\delta^{(m}_{(p}T_{q)}{}^{n)}-\tfrac13\delta^{(m}_p\delta^{n)}_qN\;,\nn\\
[\Tn1^{mn},\Tn1^{pq}]&=\epsilon^{mpr}\Tn2_r{}^{nq}|_{(mn)(pq)}\;,
\qquad  \Tn2_r{}^{mp}=\epsilon_{rnq}[\Tn1^{mn},\Tn1^{pq}]\;,    \nn\\
[\Tn1^{(mn},\Tn2_r{}^{pq)}]&=0\;,\nn\\
[\Tn1^{mn},\Tn2_r{}^{pq}]&=\epsilon^{mps}\Tn3_{sr}{}^{nq}+\epsilon^{mps}(\delta^n_r\Tn3_s{}^q-\tfrac14\delta^q_r\Tn3_s{}^n)\,|_{(mn)(pq)}\;,\\ 
[\Tn2_p{}^{mn},\Tn{-1}_{qr}]&=2\epsilon_{ps(q}\delta_{r)}^{(m}\Tn1^{n)s}\;,\nn\\
[\Tn{-1}_{mn},\Tn3_{pq}{}^{rs}]&=-2\epsilon_{mpt}(\delta^r_n\Tn2_q{}^{st}-\tfrac15\delta^r_q\Tn2_n{}^{st})|_{(mn)(pq)(rs)}\;,\nn\\
[\Tn{-1}_{mn},\Tn3_p{}^q]&=-\tfrac85\epsilon_{ps(m}\Tn2_{n)}{}^{qs}\;.\nn
\end{align}
The relative coeffients in the first equation are determined by the ideal generated by $[\Tn1^{(mn},\Tn2_r{}^{pq)}]$ 
at level~3.
Furthermore,
\begin{align}
\Tn1^{mn}\cdot \En1_p&=\delta^{(m}_p\En2^{n)}\;,\nn\\
\Tn{-1}_{mn}\cdot \En2^{p}&=\delta^p_{(m}\En1_{n)}\;,\\
\Tn1^{mn}\cdot \En2^{p}&=\En3^{mnp}+\epsilon^{pq(m}\En3_q{}^{n)}\;,\nn\\
\Tn2_p{}^{mn}\cdot \En1_q&=-\epsilon_{pqr}\En3^{mnr}
		+2(\delta^{(m}_q\En3_p{}^{n)}-\tfrac14\delta^{(m}_p\En3_q{}^{n)})\;.\nn
\end{align}

For the brackets involving $J$, start the cocycle by 
\begin{align}
[\Tn{-1}_{mn},\Jn1_p]=\epsilon_{pq(m}T_{n)}{}^q\;.
\end{align}
Then, since necessarily $[\Tn{-1}_{mn},\Jn2^p]=\delta_{(m}^p\Jn1_{n)}$, the Jacobi identity
$[\Tn1,\Tn{-1},\Jn1]$ demands
\begin{align}
[\Tn1^{mn},\Jn1_p]&=\Tn2_p{}^{mn}+\delta_p^{(m}\Jn2^{n)}\;.
\end{align}

The  solution of the linear cocycle up to level 3 shows an arbitrariness due to the common module $(11)$ in $\Tn3$ and $\Jn3$, and the possible redefinition of
$\Jn3_m{}^n$ by $\Jn3_m{}^n\rightarrow\Jn3_m{}^n+\alpha\Tn3_m{}^n$. Demanding that $[\Jn1,\Jn2]=\Jn3$ fixes this arbitrariness, so that all remaining redefinitions are rescalings. The solution to all relevant Jacobi identities is, up to level~3:
\begin{align}
[\Tn{-1}_{mn},\Jn1_p]&=\epsilon_{pq(m}T_{n)}{}^q\;,\nn\\
[\Tn{-1}_{mn},\Jn2^p]&=\delta_{(m}^p\Jn1_{n)}\;,\nn\\
[\Tn{-1}_{mn},\Jn3^{pqr}]&=\delta^{(p}_{(m}\Tn2_{n)}{}^{qr)}+\delta^{(p}_m\delta^q_n\Jn2^{r)}\;,\nn\\
[\Tn{-1}_{mn},\Jn3_p{}^q]&=-\tfrac23\delta^q_{(m}\epsilon_{n)ps}\Jn2^s\;,\\
[\Tn1^{mn},\Jn1_p]&=\Tn2_p{}^{mn}+\delta_p^{(m}\Jn2^{n)}\;.\nn\\
[\Tn1^{mn},\Jn2^p]&=-\tfrac5{12}\epsilon^{pq(m}\Tn3_q{}^{n)}+\epsilon^{pq(m}\Jn3_q{}^{n)}+\Jn3^{mnp}\;,\nn\\
[\Tn2_p{}^{mn},\Jn1_q]&=2\Tn3_{pq}{}^{mn}+\tfrac23(\delta^{(m}_q\Tn3_p{}^{n)}-\tfrac14\delta^{(m}_p\Tn3_q{}^{n)})\nn\\
                                   &\qquad+2(\delta^{(m}_q\Jn3_p{}^{n)}-\tfrac14\delta^{(m}_p\Jn3_q{}^{n)})-\epsilon_{pqr}\Jn3^{mnr}\;\nn
\end{align}
and
\begin{align}
[\Jn1_m,\Jn1_n]&=-2\epsilon_{mnp}\Jn2^p\;,\nn\\
[\Jn1_m,\Jn2^n]&=-\tfrac92\Jn3_m{}^n\;.
\end{align}
We note the absence of $\Tn2$ in $[\Tn{-1},\Jn3_{(11)}]$, which arises when $[\Jn1,\Jn2]=\Jn3$ is demanded.

The action of $J$ can of course be extended to $\Tn{n}$, $n<0$, by using the Leibniz property (Jacobi identity)
$[J,[A,B]]=[[J,A],B]+[A,[J,B]]$ for the generation of $(\fg^{++})_{<0}$ by $\Tn{-1}$. For example,
\begin{align}
[\Jn1_m,\Tn{-2}_{np}{}^q]&=4(\delta^q_m\Tn{-1}_{np}-\tfrac12\delta^q_{(n}\Tn{-1}_{p)m})\;,\nn\\
[\Jn2^m,\Tn{-2}_{np}{}^q]&=2(\delta^m_{(n}T_{p)}{}^q-\tfrac14\delta^q_{(n}T_{p)}{}^m)\;,
\end{align}
etcetera.

Focussing on the action of $\Jn1$ and specifically its part mapping $\fg^{++}\rightarrow\fg^{++}$ (the coefficients $u_{m\alpha}{}^\beta$),
the following mappings are obtained in the ``central'' part of $\fg^{++}$.
\begin{equation}
\begin{tikzcd}[row sep = -10 pt, column sep = 12 pt]
(20)\ar[dddr]&&&&&&&&(02)\\
&&&&&&&&\phantom{(00)}\\
(01)\ar[dr]&&&&&&&&(10)\\
&(11)\ar[dr]&&&(11)&&&(11)\ar[uuur]\ar[ur]\ar[dr]\\
2(12)\ar[ur]\ar[dr]&&(21)\ar[r]&(20)\ar[ur]&&(02)\ar[r]&(12)\ar[ur]\ar[dr]&&2(21)\\
&(22)\ar[ur]&&&(00)&&&(22)\ar[ur]\ar[dr]\ar[dddr]\\
(31)\ar[ur]&&&&&&&&(13)\\
&&&&&&&&\phantom{(00)}\\
(23)\ar[uuur]&&&&&&&&(32)
\end{tikzcd}
\label{A1++gradientPicture}
\end{equation}
This is the action on generators of $\fg^{++}$. 
Note that although an arrow $(11)\rightarrow(02)$ between degrees 0 and 1 is allowed by tensor product of representations, it is absent since all appearing modules are modules of the degree 0 subalgebra, and thus $[T,\Jn1]=\Jn1$.
The action on components of $A=A^\alpha T_\alpha\in\fg^{++}$ is obtained by replacing the modules with their duals and reversing all arrows. 
 We remind that when the transformation of $\hat\Delta_M$ as the fundamental is taken into account, all such ``derivatives'' commute. The left part of 
 eq. \eqref{A1++gradientPicture} gives gradient structures emanating from level 0 (identified with ``constant modes''). 
 
 The corresponding picture for the fundamental module of $\mA$ is given below. It relies on representation matrices $j_{MN}{}^P$, the form of which have not been given explicitly above.
 \begin{equation}
\begin{tikzcd}[row sep = -10 pt, column sep = 12 pt]
&&&2(21)\\
&&&\phantom{(00)}\\
&&(11)\ar[uur]\ar[r]\ar[ddr]&(10)\\
(10)\ar[r]&(01)\ar[ur]&&\\
&&(03)\ar[r]\ar[ddr]&(02)\\
&&&\phantom{(00)}\\
&&&(13)
\end{tikzcd}
\end{equation}

It is interesting to note that in this example, as well as in all $A_r^{++}$ and $E_{10}$ (maybe always?),
when we restrict to $\hat\Delta_m$ in the section defined by level 1, the representation matrices $j_{(mn)}{}^M$ are vanishing.
For $A_r^{++}$, levels 1 and 2 of $R(-\Lambda_{-1})$ are
$(10\ldots0)$ and $(0010\ldots01)\oplus(010\ldots0)$, respectively. Level 2 can only be obtained as an {\it antisymmetric} product of level 1, where it indeed appears as $[\Jn1_m,\Jn1_n]=\Jn2_{mn}$. 
The $E_{10}$ fundamental starts at level 3 in the grading with respect to the exceptional node. Levels 3 and 6 are identical to levels 1 and 2 in the $A_8^{++}$ fundamental module.
Then, acting in any module, successive symmetrised action of $\hat\Delta_m$ is only given by symmetrised products of representation matrices of the subalgebra $F\subset\mA$, in the case of the adjoint the matrices $u_m$, without the extra terms with $j$ appearing \eg\ in eq. \eqref{SectionQuestionEq}. 


\section{Conclusions and outlook\label{OutlookSection}}

We have demonstrated how gradient structures in over-extended KM algebras (and their modules) are obtained through the action of the extending generators $J_M$ in $R(-\Lambda_{-1})$. As far as we can judge, such a construction does not imply the (weak) section constraint, although the question deserves further investigation.
Gradients are obtained by choosing a subset of generators $J_m$ spanning a section, which are vectors under a 
$\gl(d)$ acting on the section.

It seems plausible that this observation will be a key to the connection between an extended geometry based on an over-extended algebra and the dynamics of cosmological billiards. The extended geometry should then be based on the extended algebra $\mA$, in accordance with its appearance in the relevant tensor hierarchy algebra. With gradient structures present in the algebra, we expect that there will be possible gauge choices that identify them with actual derivatives. In such 
a gauge, all space-dependence would come through level expansions in $\fg^{++}$, and space would in this sense be algebraically emergent.

Are there more extensions of over-extended KM algebras? We would not exclude this, but they do not seem to occur in tensor hierarchy algebras.  The ones treated here generalise $L_1$ for affine KM algebras. What about generalising $L_m$? It would fit as lowest state in 
$R(-m\lambda)$, but other lowest weight modules may also be candidates. In the absence of guidance from tensor hierarchy algebras, we do not know how to calculate the appropriate Lie algebra cohomologies.
In case of extensions of affine KM algebras by Virasoro generators, the Sugawara construction provides the additional generators.
Is there a similar/generalising construction for over-extended KM algebras (even for the fundamental alone)?
In refs. \cite{Damour:2007dt,Damour:2009ww}, signs of such a construction were found. We have not investigated this, and presently have nothing to add on the issue.

Another potentially interesting issue is the r\^ole of hyperbolicity. All physically relevant over-extended algebras are hyperbolic: this applies to
$A_r^{++}$ for $r\leq8$ as well as to $E_{10}$. In the context of billiards, hyperbolicity implies chaotic motion \cite{Damour:2001sa}.
In terms of constructing representatives of the Lie algebra cohomology of the extension in different gradings in Section \ref{CovDecSec}, we showed examples and argued in general that moving between gradings with respect to adjacent nodes, there must be a change of representative changing the covariance. This may fail if the algebra is not hyperbolic; if deleting a certain node does not yield a finite-dimensional or affine KM algebra, there is no clear reason why the cocycles should be tensorial with respect to such a subalgebra. 

We would finally like to comment on the continuation to very extended KM algebras $\fg^{+++}$ and the ``$E_{11}$ proposal''
\cite{West:2001as,West:2017vhm}. It is known that the module now extending $\fg^{+++}$ still is $R(-\Lambda_{-1})$, \ie, associated to the second node from the left of the very extended Dynkin diagram (together with more modules) 
\cite{Bossard:2017wxl,Bossard:2021ebg}.
In a chosen section, $R(-\Lambda_{-1})$ starts out with an antisymmetric bivector rather than a vector, and it already seems that the opportunity to obtain derivatives algebraically has passed (setting one of the two indices as a time direction of course still gives the spatial gradient structures in a $\fg^{++}$-covariant grading), and that it is unique for over-extended KM algebras.


\begin{appendix}

\section{Untwisted affine Kac--Moody algebras and representations\label{AffineAppendix}}

Given a finite-dimensional simple Lie algebra $\fg$ with a basis $\{T_a\}$, its untwisted affine extension $\fg^+$ is spanned by the loop generators
$\{T_{a,m}\}_{m\in\ZZ}$ and the central element $\KK$, with Lie brackets
\begin{align}
[T_{a,m},T_{b,n}]=f_{ab}{}^cT_{c,m+n}+m\delta_{m+n,0}\eta_{ab}\KK\;,
\end{align}
where $f$ and $\eta$ are the structure constants and Cartan--Killing metric of $\fg$.
The Cartan matrix is degenerate, in order to cure this, one can complement the algebra with $L_0$ (this is what happens in the further over-extended algebra, but any $L_m$ can be used).
Affine decompositions of hyperbolic KM algebras appear \eg\ in refs. \cite{Feingold:1983,Kac1988}.

At $k=\pm1$ there are highest/lowest weight modules ($k$ is the eigenvalue of $\KK$).
We focus on the ones being $R(\pm\Lambda_0)$, which are the ones relevant for the further over-extension (there may be others, in $A_r^+$ there are $r+1$ choices related to the $\ZZ_{r+1}$ symmetry of the Dynkin diagram, in $E_6^+$ there are three, while in $E_9$ there is only one).

Let $f$ be the affine module $f=R(-\Lambda_0)$ at level 1 ($k=-1$). We denote basis vectors $E_\mu$, and have representation matrices
$(t_{a,m})_\mu{}^\nu$, and also Virasoro representation matrices $(\ell_m)_\mu{}^\nu$. 

The Virasoro generators at definite $k$ may be obtained through the Sugawara construction \cite{Sugawara:1967rw} as
\begin{align}
L_m^{(k)}={1\over2(g^\vee+k)}\sum_{n\in\ZZ}\eta^{ab}:T_{a,n}T_{b,m-n}:\;.
\end{align}
The central charge is $c^{(k)}={k\dim\fg\over g^\vee+k}$ ($g^\vee$ is the dual Coxeter number of $\fg$).

At level 2, there are invariant tensors.
The affine invariant tensors $C_m\in\End(\otimes^2R(\lambda))$ are defined as
\begin{align}
C_m=\ell_m\otimes1+1\otimes\ell_m-\sum_{n\in\ZZ}\eta^{ab}t_{a,n}\otimes t_{b,m-n}\;.
\end{align}
They are $g^\vee+2$ times the coset Virasoro operators $L^{(\mathrm{coset})}_m=\ell_m\otimes1+1\otimes\ell_m-L_m^{(k=2)}$, with central charge $c^{(\mathrm{coset})}={2\dim\fg\over(g^\vee+1)(g^\vee+2)}$.

The tensor product $\otimes^2f$ contains an infinite number of irreducible lowest weight
modules, organised in a (finite) number of modules of the coset Virasoro algebra. The details of these modules of course depend on the central charge.
Irrespective of the central charge, the leading symmetric and antisymmetric modules are annihilated by $C_0$ and $C_0-2$ respectively
(these are the symmetric and antisymmetric part of the $Y$ tensor of extended geometry 
\cite{Berman:2012vc,Bossard:2017aae,Cederwall:2017fjm}).

\section{Affine decomposition of over-extended Kac--Moody alge\-bras and representations\label{AffineDecompSec}}

The grading operator with respect to the over-extending node (number $-1$ in the Dynkin diagrams) is $-\KK$.
The remaining Cartan generator (not in the Cartan subalgebra of $\fg$) is $\dd$, acting as $L_0$ on the affine subalgebra at $k=0$, $[\dd,T_{a,m}]=-mT_{a,m}$.
At level $\pm1$ the generators are in the lowest/highest weight fundamental, but with a shifted action of $\dd$,
$[\dd,T_\mu]=-(\ell_0-1)_\mu{}^\nu T_\nu$, $[\dd,\bar T^\mu]=(\ell_0-1)_\nu{}^\mu \bar T^\nu$, where $\ell_0$ is the representation matrix for $L_0$ at $k=\pm1$.
The remaining brackets in the local Lie algebra are
\begin{align}
[T_\mu,\bar T^\nu]=\delta_\mu^\nu\dd+(\ell_0-1)_\mu{}^\nu\KK-\sum_{m\in\ZZ}\eta^{ab}(t_{a,m})_\mu{}^\nu T_{b,-m}\;.
\end{align}

At level 2, there is $[T_\mu,T_\nu]$. Calculating
\begin{align}
[[T_\mu,T_\nu],\bar T_\kappa]=-2(C_0-2)_{[\mu\nu]}{}^{\kappa\lambda}T_\lambda\;
\end{align}
tells that the leading antisymmetric module generates an ideal, leaving the generators 
$T_{\mu\nu}=\tfrac12[T_\mu,T_\nu]$ in all subleading antisymmetric modules at level 2.

We thus have $(\fg^{++})_0=\langle\dd\rangle\lplus\fg^+$, $(\fg^{++})_1=f[1]$, $(\fg^{++})_2=a_2[2]$, where shift refers to mode number,
$f$ is the affine lowest weight fundamental and $a_2$ contains all subleading module in $\wedge^2f$.

The fundamental lowest weight module of $\fg^{++}$, $R(-\Lambda_{-1})$ contains at level 0 only the lowest vector
$E\in\CC[-1]$ with $\bar T^\mu\cdot E=0$, $T_{a,m}\cdot E=0$, $\dd\cdot E=-E$. At level 1 there is $E_\mu=T_\mu\cdot E\in f$, with
$\dd\cdot E_\mu=-(\ell_0)_\mu{}^\nu E_\nu$, $\bar T^\mu\cdot E_\nu=\delta^\mu_\nu E$.
 At level 2, let $E_{\mu\nu}=T_\mu\cdot T_\nu\cdot E$. Then,
\begin{align}
\bar T^\kappa\cdot E_{\mu\nu}=(C_0-1+\sigma)_{\mu\nu}{}^{\kappa\lambda}E_\lambda\;,
\end{align}
where $\sigma_{\mu\nu}{}^{\kappa\lambda}=\delta_\mu^\lambda\delta_\nu^\kappa$ is the permutation operator.
Leading symmetric and antisymmetric parts of $(\otimes^2 f)[1]$ are annihilated, and we are left with
$(s_2+a_2)[1]$ at level 2, where $s_2$ consists of all subleading affine modules in $\vee^2f$.

The symmetric and antisymmetric products $\vee^2F$ and $\wedge^2F$ are, up to level 2,
\begin{align}
\vee^2F&=\CC[-2]\oplus f[-1]\oplus s_2^\compl\oplus 2s_2\oplus a_2\oplus\ldots \;,\nn\\
\wedge^2F&=f[-1]\oplus s_2\oplus a_2^\compl\oplus2a_2\oplus\ldots\;,
\end{align} 
where $s_2^\compl$ and $a_ 2^\compl$ are the complements to $s_2$ and $a_2$ in $\vee^2f$ and $\wedge^2f$, respectively, \ie, the leading symmetric and antisymmetric affine modules.
Comparing to the leading symmetric and antisymmetric $\fg^{++}$ modules, which are
\begin{align}
R(-2\Lambda_{-1})&=\CC[-2]\oplus f[-1]\oplus s_2^\compl\oplus s_2\oplus a_2\oplus\ldots \;,\nn\\
R(-2\Lambda_{-1}+\alpha_{-1})&=f[-1]\oplus s_2\oplus a_2^\compl\oplus a_2\oplus\ldots\;,
\end{align}
we find subleading symmetric and antisymmetric $\fg^{++}$ modules starting with $s_2$ and $a_2$ at level 2. These level 2 states can be constructed explicitly by using that they must be annihilated by $\bar T^\mu$; they are
\begin{align}
&E\otimes E_{(\mu\nu)}+E_{(\mu\nu)}\otimes E-(C_0)_{(\mu\nu)}{}^{\kappa\lambda}E_\kappa\otimes E_\lambda\nn\\
\hbox{and}\qquad
&E\otimes E_{[\mu\nu]}-E_{[\mu\nu]}\otimes E-(C_0-2)_{[\mu\nu]}{}^{\kappa\lambda}E_\kappa\otimes E_\lambda\;.
\label{LeadingSubleading}
\end{align}

\section{Conjectures vs. proofs\label{ConjProofAppendix}}

Over-extended KM algebras and their representations are complicated to deal with. There are of course Weyl--Kac denominator and character formulas \cite{Kac}, but the sums over the Weyl group do not result in closed expressions in terms of known functions.
We deal with structure constants and representation matrices for over-extended KM algebras as if they are ``known''.

Essential parts of the present work rely on conjectures, which have strong support in many examples, but for which general proofs are lacking.
For clarity, we would like to point them out.

One conjecture which has not been proven, and which is heavily used in the present paper, is the structure of the tensor hierarchy algebra extension $S(\fg^{++},\lambda)$ of an over-extended KM algebra, referred to in Section \ref{IntroSec}.
The form (as a vector space) of a Borcherds superalgebra and a shifted copy of the same Borcherds superalgebra implies the existence of the cocycles relevant for the extension/deformation in Section \ref{FundExtSec}.
The evidence for the conjecture in ref. \cite{Cederwall:2021ymp} is abundant. A strict proof is desirable.

A second property of the extended algebra which is recurringly used is the statement that it is possible to find a representative for the Lie algebra cohomology so that the brackets $[J,J]$ contains no part in $\fg^{++}$ and thus forms a subalgebra. In ref. \cite{Cederwall:2021ymp} we tried to sketch a proof using the grading by $\fg^+$ modules of Section \ref{AffineGradingSection} and Appendix \ref{AffineDecompSec}. Again, having a strict proof would be desirable.

\end{appendix}

\subsection*{Acknowledgements}

Part of the research presented was conducted during the program ``Cohomological Aspects of Quantum Field Theory'' at the Mittag-Leffler Institute, Djursholm, Sweden, supported by the Swedish Research Council under grant no. 2021-06594.

\bibliographystyle{utphysmod2}


\begin{thebibliography}{10}

\bibitem{Cederwall:2021ymp}
M.~Cederwall and J.~Palmkvist,  {\em {Tensor hierarchy algebra extensions of
  over-extended Kac--Moody algebras}}, {Commun. Math. Phys.} ({2021})
  [\href{http://www.arXiv.org/abs/2103.02476}{{\tt 2103.02476}}].

\bibitem{Ehlers:1957zz}
J.~Ehlers,  {\em {Konstruktionen und Charakterisierung von L\"osungen der
  Einsteinschen Gravitationsfeldgleichungen}},. PhD thesis, Hamburg, 1957.

\bibitem{Geroch:1972yt}
R.~P. Geroch,  {\em {A method for generating new solutions of Einstein's
  equation. 2}}, J. Math. Phys. {\bf 13}, 394--404 (1972).

\bibitem{BKL}
V.~A. Belinskii, I.~M. Khalatnikov and E.~M. Lifshitz,  {\em Oscillatory
  approach to a singular point in the relativistic cosmology}, Adv. Phys. {\bf
  19}, 525 (1970).

\bibitem{Damour:2001sa}
T.~Damour, M.~Henneaux, B.~Julia and H.~Nicolai,  {\em {Hyperbolic Kac--Moody
  algebras and chaos in Kaluza--Klein models}}, Phys. Lett. B {\bf 509},
  323--330 (2001) [\href{http://www.arXiv.org/abs/hep-th/0103094}{{\tt
  hep-th/0103094}}].

\bibitem{Damour:2002mp}
T.~Damour, M.~Henneaux and H.~Nicolai,  {\em {Billiard dynamics of
  Einstein-matter systems near a spacelike singularity}}, in {\em {School on
  Quantum Gravity}}, pp.~207--247.
\newblock 1, 2002.

\bibitem{Damour:2002et}
T.~Damour, M.~Henneaux and H.~Nicolai,  {\em {Cosmological billiards}}, Class.
  Quant. Grav. {\bf 20}, R145--R200 (2003)
  [\href{http://www.arXiv.org/abs/hep-th/0212256}{{\tt hep-th/0212256}}].

\bibitem{Henneaux:2007ej}
M.~Henneaux, D.~Persson and P.~Spindel,  {\em {Spacelike singularities and
  hidden symmetries of gravity}}, Living Rev. Rel. {\bf 11}, 1 (2008)
  [\href{http://www.arXiv.org/abs/0710.1818}{{\tt 0710.1818}}].

\bibitem{Damour:2002cu}
T.~Damour, M.~Henneaux and H.~Nicolai,  {\em {E(10) and a 'small tension
  expansion' of M theory}}, Phys. Rev. Lett. {\bf 89}, 221601 (2002)
[\href{http://www.arXiv.org/abs/hep-th/0207267}{{\tt hep-th/0207267}}].

\bibitem{Kleinschmidt:2005bq}
A.~Kleinschmidt and H.~Nicolai,  {\em {Gradient representations and affine
  structures in AE(n)}}, Class. Quant. Grav. {\bf 22}, 4457--4488 (2005)
  [\href{http://www.arXiv.org/abs/hep-th/0506238}{{\tt hep-th/0506238}}].

\bibitem{Cederwall:2017fjm}
M.~Cederwall and J.~Palmkvist,  {\em {Extended geometries}}, JHEP {\bf 02}, 071
  (2018)
[\href{http://www.arXiv.org/abs/1711.07694}{{\tt 1711.07694}}].

\bibitem{Cederwall:2018aab}
M.~Cederwall and J.~Palmkvist,  {\em {$L_{\infty }$ algebras for extended
  geometry from Borcherds superalgebras}}, Commun. Math. Phys. {\bf 369},
  721--760 (2019)
[\href{http://www.arXiv.org/abs/1804.04377}{{\tt 1804.04377}}].

\bibitem{Cederwall:2019qnw}
M.~Cederwall and J.~Palmkvist,  {\em {Tensor hierarchy algebras and extended
  geometry. Part I. Construction of the algebra}}, JHEP {\bf 02}, 144 (2020)
  [\href{http://www.arXiv.org/abs/1908.08695}{{\tt 1908.08695}}].

\bibitem{Cederwall:2019bai}
M.~Cederwall and J.~Palmkvist,  {\em {Tensor hierarchy algebras and extended
  geometry. Part II. Gauge structure and dynamics}}, JHEP {\bf 02}, 145 (2020)
  [\href{http://www.arXiv.org/abs/1908.08696}{{\tt 1908.08696}}].

\bibitem{Cederwall:2021xqi}
M.~Cederwall and J.~Palmkvist,  {\em {Teleparallelism in the algebraic approach
  to extended geometry}}, JHEP {\bf 04}, 164 (2022)
  [\href{http://www.arXiv.org/abs/2112.08403}{{\tt 2112.08403}}].

\bibitem{Cederwall:2023xbj}
M.~Cederwall and J.~Palmkvist,  {\em {The teleparallel complex}}, JHEP {\bf
  05}, 068 (2023) [\href{http://www.arXiv.org/abs/2303.15391}{{\tt
  2303.15391}}].

\bibitem{Bossard:2024gry}
G.~Bossard, M.~Cederwall and J.~Palmkvist,  {\em {Teleparallel Geroch
  geometry}}, JHEP {\bf 08}, 076 (2024)
  [\href{http://www.arXiv.org/abs/2402.04055}{{\tt 2402.04055}}].

\bibitem{Tseytlin:1990va}
A.~A. Tseytlin,  {\em {Duality symmetric closed string theory and interacting
  chiral scalars}}, Nucl. Phys. {\bf B350}, 395--440
(1991).

\bibitem{Siegel:1993xq}
W.~Siegel,  {\em {Two vierbein formalism for string inspired axionic gravity}},
  Phys. Rev. {\bf D47}, 5453--5459 (1993)
[\href{http://www.arXiv.org/abs/hep-th/9302036}{{\tt hep-th/9302036}}].

\bibitem{Siegel:1993bj}
W.~Siegel,  {\em {Manifest duality in low-energy superstrings}}, in {\em
  {International Conference on Strings 93 Berkeley, California, May 24-29,
  1993}}, pp.~353--363.
\newblock 1993.
\newblock
\href{http://www.arXiv.org/abs/hep-th/9308133}{{\tt hep-th/9308133}}.
\newblock

\bibitem{Hitchin:2010qz}
N.~Hitchin,  {\em {Lectures on generalized geometry}},
\href{http://www.arXiv.org/abs/1008.0973}{{\tt 1008.0973}}.

\bibitem{Hull:2004in}
C.~M. Hull,  {\em {A geometry for non-geometric string backgrounds}}, JHEP {\bf
  10}, 065 (2005)
[\href{http://www.arXiv.org/abs/hep-th/0406102}{{\tt hep-th/0406102}}].

\bibitem{Hull:2006va}
C.~M. Hull,  {\em {Doubled geometry and T-folds}}, JHEP {\bf 07}, 080 (2007)
[\href{http://www.arXiv.org/abs/hep-th/0605149}{{\tt hep-th/0605149}}].

\bibitem{Hull:2009mi}
C.~M. Hull and B.~Zwiebach,  {\em {Double field theory}}, JHEP {\bf 09}, 099
  (2009)
[\href{http://www.arXiv.org/abs/0904.4664}{{\tt 0904.4664}}].

\bibitem{Hohm:2010jy}
O.~Hohm, C.~M. Hull and B.~Zwiebach,  {\em {Background independent action for
  double field theory}}, JHEP {\bf 07}, 016 (2010)
[\href{http://www.arXiv.org/abs/1003.5027}{{\tt 1003.5027}}].

\bibitem{Hohm:2010pp}
O.~Hohm, C.~M. Hull and B.~Zwiebach,  {\em {Generalized metric formulation of
  double field theory}}, JHEP {\bf 08}, 008 (2010)
[\href{http://www.arXiv.org/abs/1006.4823}{{\tt 1006.4823}}].

\bibitem{Jeon:2012hp}
I.~Jeon, K.~Lee, J.-H. Park and Y.~Suh,  {\em {Stringy unification of type IIA
  and IIB supergravities under $N=2$ $D=10$ supersymmetric double field
  theory}}, Phys. Lett. {\bf B723}, 245--250 (2013)
[\href{http://www.arXiv.org/abs/1210.5078}{{\tt 1210.5078}}].

\bibitem{Park:2013mpa}
J.-H. Park,  {\em {Comments on double field theory and diffeomorphisms}}, JHEP
  {\bf 06}, 098 (2013)
[\href{http://www.arXiv.org/abs/1304.5946}{{\tt 1304.5946}}].

\bibitem{Berman:2014jba}
D.~S. Berman, M.~Cederwall and M.~J. Perry,  {\em {Global aspects of double
  geometry}}, JHEP {\bf 09}, 066 (2014)
[\href{http://www.arXiv.org/abs/1401.1311}{{\tt 1401.1311}}].

\bibitem{Cederwall:2014kxa}
M.~Cederwall,  {\em {The geometry behind double geometry}}, JHEP {\bf 09}, 070
  (2014)
[\href{http://www.arXiv.org/abs/1402.2513}{{\tt 1402.2513}}].

\bibitem{Cederwall:2014opa}
M.~Cederwall,  {\em {T-duality and non-geometric solutions from double
  geometry}}, Fortsch. Phys. {\bf 62}, 942--949 (2014)
[\href{http://www.arXiv.org/abs/1409.4463}{{\tt 1409.4463}}].

\bibitem{Cederwall:2016ukd}
M.~Cederwall,  {\em {Double supergeometry}}, JHEP {\bf 06}, 155 (2016)
[\href{http://www.arXiv.org/abs/1603.04684}{{\tt 1603.04684}}].

\bibitem{Hull:2007zu}
C.~M. Hull,  {\em Generalised geometry for {M}-theory}, JHEP {\bf 0707}, 079
  (2007)
[\href{http://www.arXiv.org/abs/hep-th/0701203}{{\tt hep-th/0701203}}].

\bibitem{Pacheco:2008ps}
P.~P. Pacheco and D.~Waldram,  {\em {M}-theory, exceptional generalised
  geometry and superpotentials}, JHEP {\bf 0809}, 123 (2008)
[\href{http://www.arXiv.org/abs/0804.1362}{{\tt 0804.1362}}].

\bibitem{Hillmann:2009pp}
C.~Hillmann,  {\em {$E_{7(7)}$ and $d=11$ supergravity}},
  \href{http://www.arXiv.org/abs/0902.1509}{{\tt 0902.1509}}.
PhD thesis, Humboldt-Universit\"at zu Berlin, 2008.

\bibitem{Berman:2010is}
D.~S. Berman and M.~J. Perry,  {\em Generalized geometry and {M} theory}, JHEP
  {\bf 1106}, 074 (2011)
[\href{http://www.arXiv.org/abs/1008.1763}{{\tt 1008.1763}}].

\bibitem{Berman:2011pe}
D.~S. Berman, H.~Godazgar and M.~J. Perry,  {\em {$SO(5,5)$} duality in
  {M}-theory and generalized geometry}, Phys.Lett. {\bf B700}, 65--67 (2011)
[\href{http://www.arXiv.org/abs/1103.5733}{{\tt 1103.5733}}].

\bibitem{Coimbra:2011ky}
A.~Coimbra, C.~Strickland-Constable and D.~Waldram,  {\em {$E_{d(d)} \times
  \mathbb{R}^+$ generalised geometry, connections and M theory}}, JHEP {\bf
  1402}, 054 (2014)
[\href{http://www.arXiv.org/abs/1112.3989}{{\tt 1112.3989}}].

\bibitem{Coimbra:2012af}
A.~Coimbra, C.~Strickland-Constable and D.~Waldram,  {\em Supergravity as
  generalised geometry {II}: {$E_{d(d)} \times \mathbb{R}^+$} and {M} theory},
  JHEP {\bf 1403}, 019 (2014)
[\href{http://www.arXiv.org/abs/1212.1586}{{\tt 1212.1586}}].

\bibitem{Berman:2012vc}
D.~S. Berman, M.~Cederwall, A.~Kleinschmidt and D.~C. Thompson,  {\em {The
  gauge structure of generalised diffeomorphisms}}, JHEP {\bf 01}, 064 (2013)
[\href{http://www.arXiv.org/abs/1208.5884}{{\tt 1208.5884}}].

\bibitem{Park:2013gaj}
J.-H. Park and Y.~Suh,  {\em {U}-geometry : {SL(5)}}, JHEP {\bf 04}, 147 (2013)
[\href{http://www.arXiv.org/abs/1302.1652}{{\tt 1302.1652}}].

\bibitem{Cederwall:2013naa}
M.~Cederwall, J.~Edlund and A.~Karlsson,  {\em {Exceptional geometry and tensor
  fields}}, JHEP {\bf 07}, 028 (2013)
[\href{http://www.arXiv.org/abs/1302.6736}{{\tt 1302.6736}}].

\bibitem{Cederwall:2013oaa}
M.~Cederwall,  {\em {Non-gravitational exceptional supermultiplets}}, JHEP {\bf
  07}, 025 (2013)
[\href{http://www.arXiv.org/abs/1302.6737}{{\tt 1302.6737}}].

\bibitem{Aldazabal:2013mya}
G.~Aldazabal, M.~Gra\~{n}a, D.~Marqu\'es and J.~A. Rosabal,  {\em {Extended
  geometry and gauged maximal supergravity}}, JHEP {\bf 1306}, 046 (2013)
[\href{http://www.arXiv.org/abs/1302.5419}{{\tt 1302.5419}}].

\bibitem{Hohm:2013pua}
O.~Hohm and H.~Samtleben,  {\em Exceptional form of ${D}=11$ supergravity},
  Phys. Rev. Lett. {\bf 111}, 231601 (2013)
[\href{http://www.arXiv.org/abs/1308.1673}{{\tt 1308.1673}}].

\bibitem{Blair:2013gqa}
C.~D. Blair, E.~Malek and J.-H. Park,  {\em {M}-theory and type {IIB} from a
  duality manifest action}, JHEP {\bf 1401}, 172 (2014)
[\href{http://www.arXiv.org/abs/1311.5109}{{\tt 1311.5109}}].

\bibitem{Abzalov:2015ega}
A.~Abzalov, I.~Bakhmatov and E.~T. Musaev,  {\em {Exceptional field theory:
  $SO(5,5)$}}, JHEP {\bf 06}, 088 (2015)
  [\href{http://www.arXiv.org/abs/1504.01523}{{\tt 1504.01523}}].

\bibitem{Hohm:2013vpa}
O.~Hohm and H.~Samtleben,  {\em Exceptional field theory {I}: {E}$_{6(6)}$
  covariant form of {M}-theory and type {IIB}}, Phys. Rev. {\bf D89}, 066016
  (2014)
[\href{http://www.arXiv.org/abs/1312.0614}{{\tt 1312.0614}}].

\bibitem{Hohm:2013uia}
O.~Hohm and H.~Samtleben,  {\em Exceptional field theory {II}: {E}$_{7(7)}$},
  Phys. Rev. {\bf D89}, 066017 (2014)
[\href{http://www.arXiv.org/abs/1312.4542}{{\tt 1312.4542}}].

\bibitem{Hohm:2014fxa}
O.~Hohm and H.~Samtleben,  {\em {Exceptional field theory. III. E$_{8(8)}$}},
  Phys. Rev. {\bf D90}, 066002 (2014)
[\href{http://www.arXiv.org/abs/1406.3348}{{\tt 1406.3348}}].

\bibitem{Cederwall:2015ica}
M.~Cederwall and J.~A. Rosabal,  {\em {E$_{8}$ geometry}}, JHEP {\bf 07}, 007
  (2015)
[\href{http://www.arXiv.org/abs/1504.04843}{{\tt 1504.04843}}].

\bibitem{Bossard:2017wxl}
G.~Bossard, A.~Kleinschmidt, J.~Palmkvist, C.~N. Pope and E.~Sezgin,  {\em
  {Beyond $E_{11}$}}, JHEP {\bf 05}, 020 (2017)
[\href{http://www.arXiv.org/abs/1703.01305}{{\tt 1703.01305}}].

\bibitem{Bossard:2017aae}
G.~Bossard, M.~Cederwall, A.~Kleinschmidt, J.~Palmkvist and H.~Samtleben,  {\em
  {Generalized diffeomorphisms for $E_9$}}, Phys. Rev. {\bf D96}, 106022 (2017)
[\href{http://www.arXiv.org/abs/1708.08936}{{\tt 1708.08936}}].

\bibitem{Bossard:2018utw}
G.~Bossard, F.~Ciceri, G.~Inverso, A.~Kleinschmidt and H.~Samtleben,  {\em
  {E$_{9}$ exceptional field theory. Part I. The potential}}, JHEP {\bf 03},
  089 (2019) [\href{http://www.arXiv.org/abs/1811.04088}{{\tt 1811.04088}}].

\bibitem{Bossard:2019ksx}
G.~Bossard, A.~Kleinschmidt and E.~Sezgin,  {\em {On supersymmetric E$_{11}$
  exceptional field theory}},
\href{http://www.arXiv.org/abs/1907.02080}{{\tt 1907.02080}}.

\bibitem{Bossard:2021jix}
G.~Bossard, F.~Ciceri, G.~Inverso, A.~Kleinschmidt and H.~Samtleben,  {\em
  {E$_{9}$ exceptional field theory. Part II. The complete dynamics}}, JHEP
  {\bf 05}, 107 (2021) [\href{http://www.arXiv.org/abs/2103.12118}{{\tt
  2103.12118}}].

\bibitem{Bossard:2021ebg}
G.~Bossard, A.~Kleinschmidt and E.~Sezgin,  {\em {A master exceptional field
  theory}}, JHEP {\bf 06}, 185 (2021)
  [\href{http://www.arXiv.org/abs/2103.13411}{{\tt 2103.13411}}].

\bibitem{Bossard:2023ajq}
G.~Bossard, M.~Cederwall, A.~Kleinschmidt, J.~Palmkvist, E.~Sezgin and
  L.~Sundberg,  {\em {Extended geometry of magical supergravities}}, JHEP {\bf
  05}, 162 (2023) [\href{http://www.arXiv.org/abs/2301.10974}{{\tt
  2301.10974}}].

\bibitem{Hohm:2013jma}
O.~Hohm and H.~Samtleben,  {\em {U-duality covariant gravity}}, JHEP {\bf 09},
  080 (2013)
[\href{http://www.arXiv.org/abs/1307.0509}{{\tt 1307.0509}}].

\bibitem{Palmkvist:2013vya}
J.~Palmkvist,  {\em {The tensor hierarchy algebra}}, J. Math. Phys. {\bf 55},
  011701 (2014)
[\href{http://www.arXiv.org/abs/1305.0018}{{\tt 1305.0018}}].

\bibitem{Carbone:2018xqq}
L.~Carbone, M.~Cederwall and J.~Palmkvist,  {\em {Generators and relations for
  Lie superalgebras of Cartan type}}, J. Phys. {\bf A52}, 055203 (2019)
[\href{http://www.arXiv.org/abs/1802.05767}{{\tt 1802.05767}}].

\bibitem{Palmkvist:2015dea}
J.~Palmkvist,  {\em {Exceptional geometry and Borcherds superalgebras}}, JHEP
  {\bf 11}, 032 (2015)
[\href{http://www.arXiv.org/abs/1507.08828}{{\tt 1507.08828}}].

\bibitem{Cederwall:2015oua}
M.~Cederwall and J.~Palmkvist,  {\em {Superalgebras, constraints and partition
  functions}}, JHEP {\bf 08}, 036 (2015)
[\href{http://www.arXiv.org/abs/1503.06215}{{\tt 1503.06215}}].

\bibitem{Cederwall:2023wxc}
M.~Cederwall, S.~Jonsson, J.~Palmkvist and I.~Saberi,  {\em {Canonical
  supermultiplets and their Koszul duals}}, Commun. Math. Phys. {\bf 405}, 127
  (2024) [\href{http://www.arXiv.org/abs/2304.01258}{{\tt 2304.01258}}].

\bibitem{Damour:2007dt}
T.~Damour, A.~Kleinschmidt and H.~Nicolai,  {\em {Constraints and the E10 coset
  model}}, Class. Quant. Grav. {\bf 24}, 6097--6120 (2007)
  [\href{http://www.arXiv.org/abs/0709.2691}{{\tt 0709.2691}}].

\bibitem{Damour:2009ww}
T.~Damour, A.~Kleinschmidt and H.~Nicolai,  {\em {Sugawara-type constraints in
  hyperbolic coset models}}, Commun. Math. Phys. {\bf 302}, 755--788 (2011)
  [\href{http://www.arXiv.org/abs/0912.3491}{{\tt 0912.3491}}].

\bibitem{West:2001as}
P.~C. West,  {\em {E(11) and M theory}}, Class. Quant. Grav. {\bf 18},
  4443--4460 (2001)
[\href{http://www.arXiv.org/abs/hep-th/0104081}{{\tt hep-th/0104081}}].

\bibitem{West:2017vhm}
P.~C. West, {\em {A brief review of E theory}}, pp.~135--176.
\newblock {World Scientific}, 2017.

\bibitem{Feingold:1983}
A.~J. Feingold and I.~B. Frenkel,  {\em {A hyperbolic Kac--Moody algebra and
  the theory of Siegel modular forms of genus 2}}, Math. Ann. {\bf 263}, 87
  (1983).

\bibitem{Kac1988}
V.~G. Kac, R.~V. Moody and M.~Wakimoto, {\em On $E_{10}$}, pp.~109--128.
\newblock Springer Netherlands, 1988.

\bibitem{Sugawara:1967rw}
H.~Sugawara,  {\em {A field theory of currents}}, Phys. Rev. {\bf 170},
  1659--1662 (1968).

\bibitem{Kac}
V.~G. Kac, {\em Infinite dimensional {L}ie algebras}.
\newblock 3rd edition, Cambridge University Press, 1990.

\end{thebibliography}

\providecommand{\href}[2]{#2}\begingroup\raggedright\endgroup

\end{document}